\documentclass[aps,prd,amsmath
,eqsecnum            % Adds section numbers to equation numbers.
,preprintnumbers
,nofootinbib         % Stops RevTeX moving footnotes to the references/bibliography
 ,showpacs          % Makes the (required) PACS numbers print out
 ,showkeys          % Makes the (required) keywords print out
,twocolumn         % for two columns per page, as in the journal.  Only do this when close to submission.
]{revtex4}
\usepackage[dvips]{graphicx}% For importing graphics: .ps, .jpg, .bmp, .pcx, & others but NOT .gif .
\usepackage{epsfig}

\usepackage{amsmath}
\usepackage{amsfonts}
\usepackage{amssymb}
\usepackage{longtable}   % For tables that are too long for 1 page.

\allowdisplaybreaks[4]     % allows page breaks in the middle of an equation group.

\newcommand{\df}{\ {\overset {\rm def} =}\ }
\newcommand{\dr}[2]{\frac {{\rm d} {#1}} {{\rm d} {#2}}}
\newcommand{\pdr}[2]{\frac {\partial {#1}} {\partial {#2}}}

\newcommand{\llim}[1] {\ {\underset {#1} {\longrightarrow}}\ }

\begin{document}

\title{Geometry and topology of the quasi-plane Szekeres model}

\author{Andrzej Krasi\'nski}
\affiliation{N. Copernicus Astronomical Centre, Polish Academy of Sciences, \\
Bartycka 18, 00 716 Warszawa, Poland}
\email{akr@camk.edu.pl}

\date {}

\begin{abstract}
This paper is a revised version of arXiv:0805.0529 and {\it Phys.Rev.} {\bf
D78}, 064038 (2008), taking into account the erratum published in {\it
Phys.Rev.} {\bf D85}, 069903(E) (2012). Geometrical and topological properties
of the quasi-plane Szekeres model and of the plane symmetric dust model are
discussed. Some related comments on the quasi-hyperbolical model are made. These
properties include: (1) The pattern of expansion in the plane symmetric case,
and the Newtonian model that imitates it; (2) The possibility of toroidal
topology of the $t =$ const sections in the plane symmetric model; (3) The
absence of apparent horizons in the quasi-plane and quasi-hyperbolic models
(they are globally trapped); (4) Description of the toroidal topology in the
Szekeres coordinates; (5) Interpretation of the mass function in the quasi-plane
model.
\end{abstract}

\maketitle

\section{Motivation}

The quasi-spherical Szekeres model \cite{Szek1975a} -- \cite{Kras1997} is rather
well-understood by now. In spite of its nontrivial geometry, its basic defining
features are not too difficult to grasp intuitively. In a simple-minded way one
may say that it is obtained when the spherical symmetry orbits in the
Lema\^{\i}tre -- Tolman model \cite{PlKr2006,Kras1997} are made nonconcentric to
destroy the symmetry, but the energy-momentum tensor is still that of dust.
Recently, that model even found application to solving problems directly related
to observational cosmology \cite{Bole2006,Bole2007}. In contrast to this, the
first serious attempt to interpret the quasi-plane and quasi-hyperbolic models
\cite{HeKr2008} revealed that even the corresponding plane- and hyperbolically
symmetric models are not really understood and require more investigation. Some
properties of those models were established in Ref. \cite{HeKr2008}, the present
paper is a continuation of that research.

The aim of the present paper is to clarify some of the basic geometrical
features of the quasi-plane Szekeres model, and of the plane symmetric dust
model. The following topics are investigated here: (1) The pattern of expansion
in the plane symmetric model, and the Newtonian model that imitates it; (2) The
possibility of toroidal topology of the $t =$ const sections in the plane
symmetric model; (3) The absence of apparent horizons in the quasi-plane and
quasi-hyperbolic models (they are globally trapped); (4) Description of the
toroidal topology in the Szekeres coordinates; (5) Interpretation of the mass
function in the quasi-plane model. For the most part, the paper is devoted to
showing that the space of constant time in the plane symmetric dust models can
be interpreted as a family of flat tori, with the ones of smaller diameter
enclosed inside those of larger diameter. Such a topology explains several
properties of the models, among them the pattern of decelerated expansion and
the finiteness of the mass function. It turns out that these models are of
lesser use in astrophysical cosmology than the quasi-spherical ones. Because of
being globally trapped, they cannot be used for modeling dynamical black holes.
Because they expand by the same law as the positive-energy Lema\^{\i}tre --
Tolman model, they cannot model the formation of structures that collapse to
very dense states. They might be applicable for the description of formation of
moderate condensations, like galaxy clusters, and of voids.

Mena, Nat\'ario and Tod also considered the quasi-plane and quasi-hyper\-bolic
Szekeres models with toroidal and higher-genus topologies \cite{MNTo2008}. They
considered the matching of those solutions, with nonzero cosmological constant
(corresponding to $\Lambda > 0$ in the notation adopted here), to the plane- and
hyperbolically symmetric counterparts of the Schwarzschild solution, also
allowed to have nontrivial topologies of the symmetry orbits. However, there is
no overlap between their results and those of the present paper, as they mainly
considered the global geometry of the resulting black hole, while here the
emphasis is put on local geometry of the topologically nontrivial Szekeres
spacetime.

The present text is a corrected version of Ref. \cite{Kras2008}, taking into
account the erratum \cite{Kras2012}. The error hereby corrected was revealed by
Charles Hellaby. By this opportunity, typos and style were corrected as well.

\section{Introducing the Szekeres solutions}\label{intszek}

\setcounter{equation}{0}

The metric of the Szekeres solutions is
\begin{equation}\label{2.1}
{\rm d} s^2 = {\rm d} t^2 - {\rm e}^{2 \alpha} {\rm d} z^2- {\rm e}^{2 \beta}
\left({\rm d} x^2 + {\rm d} y^2\right),
\end{equation}
where $\alpha$ and $\beta$ are functions of $(t, x, y, z)$ to be determined from
the Einstein equations with a dust source. The coordinates of (\ref{2.1}) are
comoving so the velocity field of the dust is $u^{\mu} = {\delta^{\mu}}_0$ and
$\dot{u}^{\mu} = 0$.

There are two families of Szekeres solutions, depending on whether $\beta,_z =
0$ or $\beta,_z \neq 0$. The first family is a simultaneous generalisation of
the Friedmann and Kantowski -- Sachs \cite{KaSa1966} models. So far it has found
no useful application in astrophysical cosmology, and we shall not discuss it
here (see Ref. \cite{PlKr2006}); we shall deal only with the second family.
After the Einstein equations are solved, the metric functions in (\ref{2.1})
become
\begin{eqnarray}\label{2.2}
{\rm e}^{\beta} &=& \Phi(t, z) / {\rm e}^{\nu(z, x, y)}, \nonumber \\
{\rm e}^{\alpha} &=& h(z) \Phi(t, z) \beta,_z \equiv h(z) \left(\Phi,_z + \Phi
\nu,_z\right), \nonumber \\
{\rm e}^{- \nu} &=& A(z)\left(x^2 + y^2\right) + 2B_1(z) x + 2B_2 (z)y + C(z),\
\ \ \ \ \
\end{eqnarray}
where the function $\Phi(t, z)$ is a solution of the equation
\begin{equation}\label{2.3}
{\Phi,_t}^2 = - k(z) + \frac {2M(z)} {\Phi} + \frac 1 3 \Lambda \Phi^2;
\end{equation}
while $h(z)$, $k(z)$, $M(z)$, $A(z)$, $B_1(z)$, $B_2(z)$ and $C(z)$ are
arbitrary functions obeying
\begin{equation}\label{2.4}
g(z) \df 4 \left(AC - {B_1}^2 - {B_2}^2\right) = 1/h^2(z) + k(z).
\end{equation}
The mass-density is
\begin{equation}\label{2.5}
\kappa \rho = \frac {\left(2M {\rm e}^{3\nu}\right),_z} {{\rm e}^{2\beta}
\left({\rm e}^{\beta}\right),_z}; \qquad \kappa = 8 \pi G / c^2.
\end{equation}
In the present paper we will mostly consider the case $\Lambda = 0$.

These solutions have in general no symmetry, and acquire a 3-dimensional
symmetry group with 2-dimensional orbits when $A$, $B_1$, $B_2$ and $C$ are all
constant (that is, when $\nu,_z = 0$). The sign of $g(z)$ determines the
geometry of the surfaces $(t =$ const, $z =$ const), and the symmetry of the
limiting solution. The geometry is spherical, plane or hyperbolic when $g > 0$,
$g = 0$ or $g < 0$, respectively. With $A$, $B_1$, $B_2$ and $C$ being functions
of $z$, the surfaces $z =$ const within a single space $t =$ const may have
different geometries (i.e. they can be spheres in one part of the space and
surfaces of constant negative curvature elsewhere, the curvature being zero at
the boundary). The sign of $k(z)$ determines the type of evolution; with $k > 0
= \Lambda$ the model expands away from an initial singularity and then
recollapses to a final singularity, with $k < 0 = \Lambda$ the model is either
ever-expanding or ever-collapsing, depending on the initial conditions; $k = 0 =
\Lambda$ is the intermediate case corresponding to the 'flat' Friedmann model.

The Robertson--Walker limit follows when $z = r$, $\Phi (t,z) = r R(t)$, $k =
k_0 r^2$ where $k_0 =$ const and $B_1 = B_2 = 0$, $C = 4A = 1$. This definition
of the R--W limit includes the definition of the limiting radial coordinate (the
Szekeres model is covariant with the transformations $z = f(z')$, where $f(z')$
is an arbitrary function).

The Szekeres models are subdivided according to the sign of $g(z)$ into
quasi-spherical (with $g > 0$), quasi-plane ($g = 0$) and quasi-hyperbolic ($g <
0$).\footnote{We stress once again that {\it the same} Szekeres model may be
quasi-spherical in one part of the spacetime, and quasi-hyperbolic elsewhere,
with the boundary between these two regions being quasi-plane, see an explicit
simple example in Ref. \cite{HeKr2008}. In most of the literature published so
far, these models have been considered separately, but this was only for
purposes of systematic research.} Despite suggestions to the contrary made in
the literature, the geometry of the latter two classes has, until very recently,
not been investigated at all and is not really understood; work on their
interpretation has only been begun by Helalby and Krasinski \cite{HeKr2008}. The
sign of $g(z)$ is independent of the sign of $k(z)$, but limitations are imposed
on $k(z)$ by the signature of the spacetime: for it to be the physical $(+ - -
-)$, the function $h^2$ must be non-negative (possibly zero at isolated points,
but not on open subsets), which, via (\ref{2.4}) means that $g(z) - k(z) \geq 0$
everywhere. Thus, with $g > 0$ (in the quasi-spherical case) all three
possibilities for $k$ are allowed; with $g = 0$ only the two $k \leq 0$
evolutions are admissible, and with $g < 0$, only the $k < 0$ evolution is
allowed.

Only the quasi-spherical model is rather well investigated, and found useful
application in astrophysical cosmology. We recall now its basic properties.

It may be imagined as such a generalisation of the Lema\^i{\i}tre--Tolman (L--T)
model in which the spheres of constant mass were made non-concentric. The
functions $A(z)$, $B_1(z)$ and $B_2(z)$ determine how the center of a sphere
changes its position in a space $t =$ const when the radius of the sphere is
increased or decreased (see a discussion of this in Ref. \cite{HeKr2002}).
Still, this is a rather simple geometry because all the arbitrary functions
depend on just one variable, $z$. They give us some limited possibility to model
real structures in the Universe (see elegant examples in Refs.
\cite{Bole2006,Bole2007}), but a fully satisfactory model should involve
arbitrary functions of all three spatial variables, to allow modelling of
arbitrary structures. Such models are still nonexistent, so the Szekeres models
are so far the best devices that exist.

Often, it is more practical to reparametrise the arbitrary functions in the
Szekeres metric as follows (this parametrisation was invented by Hellaby
\cite{Hell1996b}). Even if $A = 0$ initially, a transformation of the $(x,
y)$-coordinates can restore $A \neq 0$, so we may assume $A \neq 0$ with no loss
of generality (see Ref. \cite{PlKr2006}). Then let $g \neq 0$. Writing
\begin{eqnarray}\label{2.6}
&& \left(A, B_1, B_2\right) = \frac {\sqrt{|g|}} {2 S} (1, - P, - Q),
\quad\varepsilon \df g / |g|,\ \ \ \ \  \\
&& k = - |g| \times 2 E, \quad M = |g|^{3/2} \widetilde{M}, \quad \Phi = R
\sqrt{|g|}, \nonumber
\end{eqnarray}
we can represent the metric (\ref{2.1}) as
\begin{eqnarray}\label{2.7}
&& {\rm e}^{- \nu} / \sqrt{|g|} \df {\cal E} \df \frac S 2 \left[\left(\frac {x
- P} S\right)^2 + \left(\frac {y - Q} S\right)^2 + \varepsilon\right], \nonumber
\\
&& {\rm d} s^2 = {\rm d} t^2 - \frac {\left(R,_z - R {\cal E},_z/{\cal
E}\right)^2} {\varepsilon + 2E(z)} {\rm d} z^2 - \frac {R^2} {{\cal E}^2}
\left({\rm d} x^2 + {\rm d} y^2\right).\nonumber \\
&&
\end{eqnarray}
When $g = 0$, the transition from (\ref{2.1}) to (\ref{2.7}) is $A = 1/(2S)$,
$B_1 = - P/(2S)$, $B_2 = - Q/(2S)$, $k = - 2E$, $\widetilde{M} = M$ and $\Phi =
R$. Then (\ref{2.7}) applies with $\varepsilon = 0$, and the resulting model is
quasi-plane.

For further reference, the evolution equation (\ref{2.3}), in the variables of
(\ref{2.7}) becomes
\begin{equation}\label{2.8}
{R,_t}^2 = 2E(z) + \frac {2 \widetilde {M}(z)} R + \frac 1 3 \Lambda R^2;
\end{equation}
{}From now on, we will use this representation; the tilde over $M$ will be
dropped, but it must be remembered that the $M$ in (\ref{2.8}) is not the same
as the one in (\ref{2.3}).

The representation (\ref{2.7}) makes the calculations simpler because the
arbitrary functions in it are independent (the condition (\ref{2.4}) has been
incorporated in this form). However, it obscures the fact that the cases
$\varepsilon = +1, 0, -1$ can be parts of the same spacetime.

Rotation and acceleration of the dust source are zero, the expansion is
\begin{equation}\label{2.9}
\Theta = 3 \frac {R,_t} R + \frac {R,_{tz} - R,_t R,_z/R} {R,_z - R {\cal
E},_z/{\cal E}},
\end{equation}
and the shear tensor is
\begin{eqnarray}\label{2.10}
&& {\sigma^{\alpha}}_{\beta} = \frac 1 3 \Sigma \  {\rm diag\ } (0, 2, -1, -1),
\qquad {\rm where} \qquad \nonumber \\
&& \Sigma = \frac {\Phi,_{tz} - \Phi,_t \Phi,_z / \Phi} {\Phi,_z - \Phi \nu,_z}
\equiv \frac {R,_{tz} - R,_t R,_z / R} {R,_z - R {\cal E},_z / {\cal E}}.\ \ \ \
\ \ \ \ \ \
\end{eqnarray}

Definitions of the Szekeres solutions by invariant properties can be found in
Ref. \cite{PlKr2006}.

When $\Lambda \neq 0$, the solutions of (\ref{2.8}) involve elliptic functions.
A general formal integral of (\ref{2.8}) was presented by Barrow and
Stein-Schabes \cite{BaSS1984}. Any solution of (\ref{2.8}) will contain one more
arbitrary function of $z$ that will be denoted $t_B(z)$, and will enter the
solution in the combination $(t - t_B(z))$. The instant $t = t_B(z)$ defines the
initial moment of evolution; when $\Lambda = 0$ it is necessarily a singularity
corresponding to $\Phi = 0$, and it goes over into the Big Bang singularity in
the Friedmann limit. When $t_{B,z} \neq 0$ (that is, in general) the instant of
singularity is position-dependent, as in the L--T model.

Just as in the L--T model, another singularity may occur where $\left({\rm
e}^{\beta}\right),_z = 0$ (if this equation has solutions). This is a shell
crossing, but it is qualitatively different from that in the L--T model. As can
be seen from (\ref{2.2}), in the quasi-spherical case, when a shell crossing
exists, its intersection with a $t =$ const space will be a circle, or, in
exceptional cases, a single point, not a sphere. In the quasi-spherical models
shell crossings can be avoided altogether if the arbitrary functions are chosen
appropriately, see the complete list and derivation in Ref. \cite{HeKr2002}. In
the quasi-hyperbolic models, shell crossings can be avoided in one sheet of each
hyperboloid, but are unavoidable in the other, see Ref. \cite{HeKr2008}. In in
the quasi-plane model, if the flat surfaces existing in it are interpreted as
infinite planes, shell crossings are unavoidable \cite{HeKr2008}.

Equation (\ref{2.8}) is identical with the Friedmann equation, but, just like in
the L--T limit, with $k$ and $M$ depending on $z$, each surface $z$ = const
evolves independently of the others.

The models defined by (\ref{2.1})-- (\ref{2.5}) contain 8 functions of $z$, but
only 5 of them correspond to independent physical degrees of freedom. One of the
8 functions is determined by (\ref{2.4}), $g(z)$ was made constant by the
reparametrisation (\ref{2.6}), and one can be specified by a choice of $z$, for
example by defining $z' = M$, or $M = {z'}^3 \times \{{\rm a\ constant}\}$.

A quasi-spherical Szekeres region can be matched to the Schwarzschild solution
across a $z =$ const hypersurface \cite{Bonn1976a}. The other two Szekeres
regions can be matched to the plane- and hyperbolically symmetric counterparts
of the Schwarzschild solution (see Ref. \cite{CaDe1968} for the solutions and
\cite{HeKr2008} for the matching).

In the following, we will represent the Szekeres solutions with $\beta,_z \neq
0$ in the parametrisation introduced in (\ref{2.7}). The formula for density in
these variables is
\begin{equation}\label{2.11}
\kappa \rho = \frac {2 \left(M,_z - 3 M {\cal E},_z / {\cal E}\right)} {R^2
\left(R,_z - R {\cal E},_z / {\cal E}\right)},
\end{equation}
where, let it be recalled, the $M$ above is the $\widetilde{M}$ of (\ref{2.6}).

 \section{The plane symmetric models}\label{plsymmod}

\setcounter{equation}{0}

The plane symmetric dust models (first found by Ellis \cite{Elli1967}) result
from (\ref{2.7}) when $\varepsilon = 0$ and $(P, Q, S)$ are independent of $z$.
The constant $S$ can then be scaled to 1 by appropriate redefinitions of $R$,
$E$ and $M$. Then, with constant $P$ and $Q$, the coordinate transformation
\begin{equation}\label{3.1}
x = P + \frac {2p} {p^2 + q^2}, \qquad y = Q + \frac {2q} {p^2 + q^2}
\end{equation}
changes the metric to
\begin{equation}\label{3.2}
{\rm d} s^2 = {\rm d} t^2 - \frac {{R,_z}^2} {2E(z)}  {\rm d} z^2- {R}^2
\left({\rm d} p^2 + {\rm d} q^2\right),
\end{equation}
while the energy-density simplifies to
\begin{equation}\label{3.3}
\frac {8 \pi G} {c^2} \rho = \frac {2M,_z} {R^2 R,_z}.
\end{equation}
These models are called plane symmetric because their symmetries are the same as
those of the Euclidean plane; in the coordinates of (\ref{3.2}) they are:
\begin{subequations}\label{3.4}
\begin{eqnarray}
p' &=& p + A_1, \label{3.4a} \\
q' &=& q + A_2,\label{3.4b} \\
(p', q') &=& (p \cos \alpha + q \sin \alpha, - p \sin \alpha + q \cos \alpha),\
\ \ \ \ \  \label{3.4c}
\end{eqnarray}
\end{subequations}
where $A_1$, $A_2$ and $\alpha$ are arbitrary constants -- the group parameters.

Note that eqs. (\ref{2.8}) and (\ref{3.3}) are identical to their counterparts
in the spherically symmetric models. In particular, the function $M(z)$ enters
in the same way as the active gravitational mass did in spherical models.
However, if we wish to interpret $M(z)$ as a mass contained in a volume, we
encounter a problem -- see below.

Examples of plane symmetric spaces are the Euclidean plane and the Euclidean
space $E_3$ with the metric ${\rm d} {s_3}^2 = {\rm d} x^2 + {\rm d} y^2 + {\rm
d} z^2$. However, the space of constant $t$ in (\ref{3.2}) can never become
flat; its curvature tensor is \cite{HeKr2008}:
\begin{equation}\label{3.5}
{}^3{R^{zp}}_{zp} = {}^3{R^{zq}}_{zq} = - \frac {E,_z} {R R,_z}, \qquad
{}^3{R^{pq}}_{pq}  = - \frac {2 E} {R^2}.
\end{equation}
Nevertheless, the surfaces $P_2$ of constant $t$ and $z$ in (\ref{3.2}) are
flat. Thus, there is some mystery in the geometry of the spacetimes (\ref{3.2}).
One component of the mystery is this: In the quasi-spherical case, and in the
associated spherically symmetric model, the surfaces of constant $t$ and $z$
were spheres, and $M(z)$ was a mass inside a sphere of coordinate radius $z$. In
the plane symmetric case, if the $P_2$ surfaces are infinite planes, they do not
enclose any finite volume, so where does the mass $M(z)$ reside?

With $M$ being a constant, the metric (\ref{3.2}) becomes vacuum -- the plane
symmetric analogue of the Schwarzschild spacetime.

In the quasi-spherical Szekeres, and in spherically symmetric solutions,
analogies exist between the relativistic and the Newtonian models. We will now
compare the plane symmetric model with its possible Newtonian counterparts. For
this purpose, let us note the pattern of expansion in (\ref{3.2}) and
(\ref{2.8}) with $\Lambda = 0$. When $R,_t \neq 0$, eq. (\ref{2.8}) implies
\begin{equation}\label{3.6}
R,_{tt} = - M / R^2.
\end{equation}
Note that $R,_{tt} = 0$ implies $M = 0$, which is the Min\-kow\-ski metric. Now
take a pair of dust particles, located at $(t, z_1, p_0, q_0)$ and at $(t, z_2,
p_0, q_0)$, and consider the affine distance between them:
\begin{equation}\label{3.7}
\ell_{12}(t) = \int_{z_1}^{z_2} \frac {R,_z {\rm d} z} {\sqrt{2E}}
\Longrightarrow \dr {^2\ell_{12}} {t^2} = \int_{z_1}^{z_2} \frac {R,_{ttz} {\rm
d} z} {\sqrt{2E}}.
\end{equation}
Thus, the two particles will be receding from each other (or approaching each
other if collapse is considered) with acceleration that can never be
zero.\footnote{The acceleration would be zero if $R,_{ttz} = 0$, which leads to
a contradiction in the Einstein equations.}

Take another pair of dust particles, located at $(t, z_0, p_1, q_0)$ and at $(t,
z_0, p_2, q_0)$. The distance between them, measured within the symmetry orbit,
is
\begin{eqnarray}\label{3.8}
&& \ell_{34}(t) = \int_{p_1}^{p_2} R {\rm d} p \equiv R \left(p_2 - p_1\right)
\nonumber \\
&& \Longrightarrow \dr {^2\ell_{34}} {t^2} = R,_{tt}  \left(p_2 - p_1\right),
\end{eqnarray}
i.e. the acceleration of the expansion can never vanish in this direction,
either, unless $M = 0$. The same result will follow for any direction in the
$(p, q)$ surface. Thus, the expansion or collapse in this model proceeds with
acceleration in every spatial direction.\footnote{Since $M \geq 0$, this is fact
deceleration.} We will compare this result with the Newtonian situation.

\section{A Newtonian analogue of the plane symmetric dust
spacetime}\label{newtplsym}

\setcounter{equation}{0}

At first sight, it seems that the Newtonian model analogous to the plane
symmetric dust model should be dust whose density is constant on parallel $(x,
y)$-planes, and depends only on $z$. Let us follow this idea.

If the potential is plane symmetric, then, in the adapted coordinates, it
depends only on $z$. Thus, the Poisson equation simplifies to
\begin{equation}\label{4.1}
\dr {^2V} {z^2} = 4 \pi G \rho(z).
\end{equation}
The general solution of this is
\begin{equation}\label{4.2}
V = 4 \pi G \int_{z_0}^z {\rm d} z' \int_{z_0}^{z'} {\rm d} \widetilde{z}
\rho(\widetilde{z}) + Az + B,
\end{equation}
where $A$ and $B$ are integration constants; $z_0$ is a reference value of $z$
at which we can specify an initial condition. If we wish to have $V =$ const
(i.e. zero force) when $\rho \equiv 0$, we must take $A = 0$, and then $V(z_0) =
B$.\footnote{An $A \neq 0$ would be qualitatively similar to the cosmological
constant in relativity.} The equations of motion in this potential are
\begin{equation}\label{4.3}
\dr {v^i} t = - \pdr V {x^i},
\end{equation}
where $v^i$ are components of the velocity field of matter, so
\begin{equation}\label{4.4}
\dr {v^x} t = \dr {v^y} t = 0, \qquad \dr {v^z} t = - \dr V z = - 4 \pi G
\int_{z_0}^z {\rm d} z' \rho(z').
\end{equation}
This, however, gives a pattern of expansion different from that in the
relativistic plane symmetric model. In (\ref{4.4}), expansion with acceleration
proceeds only in the $z$-direction, while in the directions orthogonal to $z$
there is no acceleration, or, in a special case, not even any expansion.
Consequently, no obvious Newtonian analogue exists for the relativistic plane
symmetric model.\footnote{Incidentally, there will be no Newtonian analogue for
the hyperbolic model, since the orbits of hyperbolic symmetry cannot be embedded
in a Euclidean space at all. They can be embedded in a flat 3-dimensional space,
but the space then must have the signature $(- + +)$.}

Equation (\ref{4.4}) shares one property with the relativistic evolution
equation (\ref{2.8}). If $\rho$ is bounded in the range of integration, then the
force that drives the motion of the fluid is finite, giving the illusion that
the potential is generated by some finite mass. However, if we wanted to
calculate $V(z)$ by summing up contributions to it from all the volume elements
of the fluid, like is done in calculating the gravitational potentials of finite
portions of matter, then the result would be an infinite value of $V$, in
consequence of the source having infinite extent in the $(x, y)$-plane. Thus, if
we want to interpret the r.h.s. in (\ref{4.3}) as being generated by a mass,
then the mass that drives the evolution is not the total mass in the source, but
the mass of a finite portion of the source.

We now provide a solution of the Poisson equation that qualitatively mimics the
pattern of expansion of the plane symmetric relativistic model. Its
equipotential surfaces will be locally plane symmetric, but their symmetries
will not be symmetries of the whole space.

\begin{figure}
 \begin{center}
 \includegraphics[scale = 0.4]{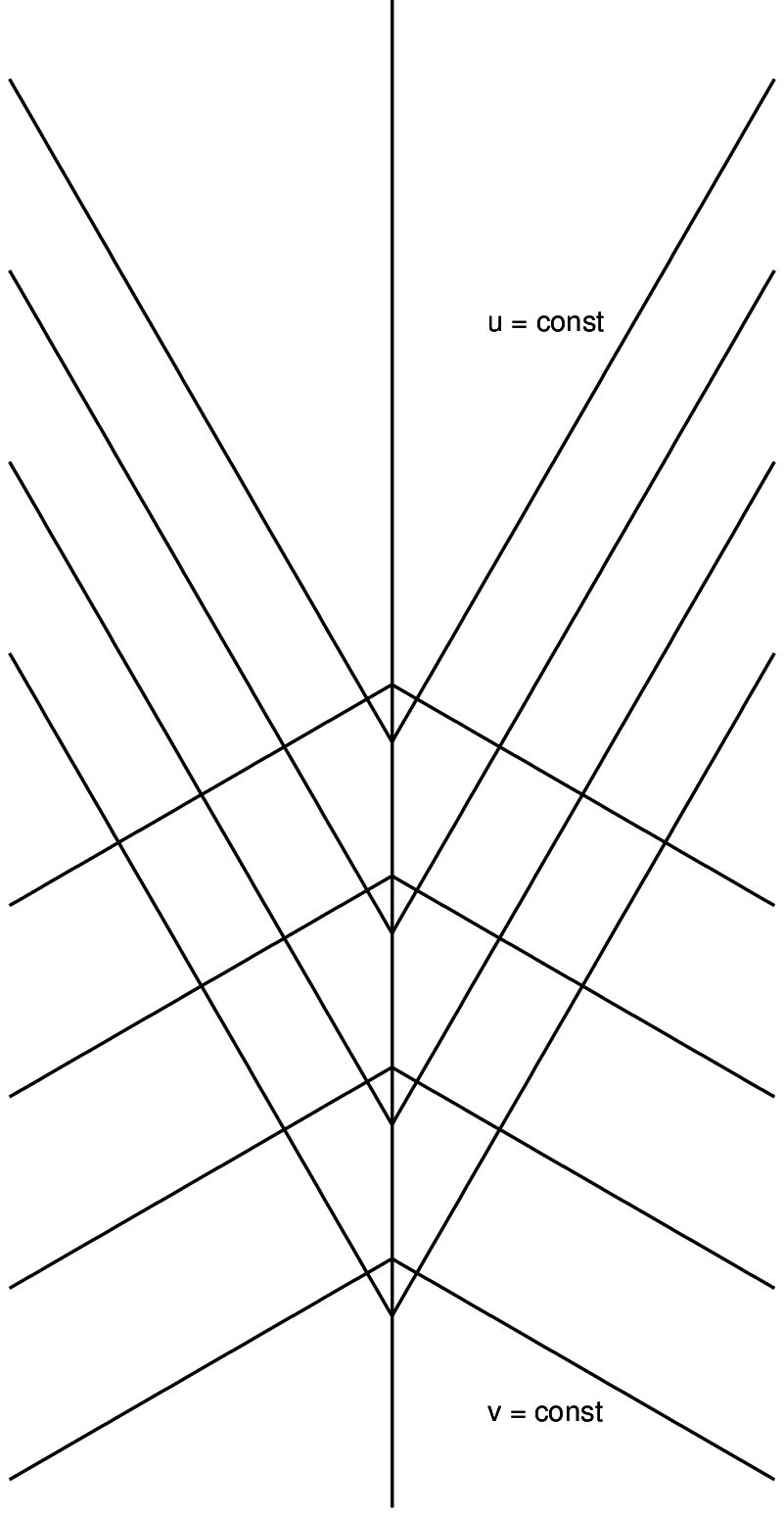}
 \caption{
 \label{coneflat}
 \footnotesize
The cones $u =$ constant are orthogonal to the cones $v =$ constant. The
coordinates in space are $u$, $v$ and the angle around the axis of symmetry. The
figure shows an axial cross-section through the setup. A gravitational potential
which depends only on $u$ in these coordinates gives an expansion pattern that
is qualitatively similar to the one in a plane symmetric dust spacetime.
 }
 \end{center}
 \end{figure}

Consider two families of cones given by the equations (see Fig. \ref{coneflat})
\begin{equation}\label{4.5}
u = z - \alpha r, \qquad v = z + r/\alpha, \qquad r \df \sqrt{x^2 + y^2},
\end{equation}
where $\alpha$ is a constant and $(x, y, z)$ are Cartesian coordinates. The
cones of constant $u$ are orthogonal to the cones of constant $v$, and the two
families are co-axial. We choose $u$ and $v$ as two coordinates in space; the
third coordinate will be the angle $\varphi$ around the axis of symmetry. We
begin with the Euclidean metric in the cylindrical coordinates, ${\rm d} s^2 =
{\rm d} r^2 + r^2 {\rm d} \varphi^2 + {\rm d} z^2$, and transform this to the
$(u, v, \varphi)$ coordinates by
\begin{equation}\label{4.6}
r = \frac {\alpha (v - u)} {1 + \alpha^2}, \qquad z = \frac {u + \alpha^2 v} {1
+ \alpha^2}.
\end{equation}
The transformed metric is
\begin{equation}\label{4.7}
{\rm d} s^2 = \frac {{\rm d} u^2 + \alpha^2 {\rm d} v^2} {1 + \alpha^2} + \frac
{\alpha^2 (v - u)^2 {\rm d} \varphi^2} {\left(1 + \alpha^2\right)^2}.
\end{equation}

The Laplace operator, which in the cylindrical coordinates is
\begin{equation}\label{4.8}
\Delta V = \frac 1 r \pdr {} r \left(r \pdr V r\right) + \frac 1 {r^2} \pdr {^2
V} {\varphi^2} + \pdr {^2 V} {z^2},
\end{equation}
in the $(u, v, \varphi)$-coordinates becomes
\begin{eqnarray}\label{4.9}
\Delta V &=& \left(1 + \alpha^2\right) \left[- \frac 1 {v - u} \pdr V u + \pdr
{^2 V} {u^2} + \frac 1 {\alpha^2 (v - u)} \pdr V v \right. \nonumber \\
&&\ \ \ + \left.\frac 1 {\alpha^2} \pdr {^2 V} {v^2} + \frac {1 + \alpha^2}
{\alpha^2 (v - u)^2} \pdr {^2 V} {\varphi^2}\right].
\end{eqnarray}
Thus, if $V$ depends only on $u$, then the Poisson equation says:
\begin{equation}\label{4.10}
\left(1 + \alpha^2\right) \left(- \frac 1 {v - u} \pdr V u + \pdr {^2 V}
{u^2}\right) = - 4 \pi G \rho.
\end{equation}

The gradient of $V(u)$ has nonzero components in all directions, and so will
create expansion decelerated in all directions. The expansion will be isotropic
with respect to the $u = v$ axis, and the anisotropy between the $(x, y)$ and
the $z$-directions is controlled by $\alpha$.

This potential was introduced here for illustrative purposes. In order to make
it credible, one should solve the continuity equation and the Euler equations of
motion in it. We do not quote here the appropriate calculations because they
lead to an intransparent tangle of differential equations. For dust, that set is
overdetermined, so probably has no solutions.

 \section{Plane symmetric 3-spaces interpreted as tori}\label{PlanSym3Sp}

\setcounter{equation}{0}

Although known for a long time (see Ref. \cite{Elli1967}), the plane symmetric
model has not been investigated for its geometrical and physical properties.

Since a flat spatial geometry is not possible in it (see eq. (\ref{3.5})), we
now consider other possible 3-geometries with planar symmetry. The next simplest
is a space of constant curvature. {}From (\ref{3.5}), the space of constant $t =
t_0$ will have constant curvature when
\begin{equation}\label{5.1}
2E = \pm C^2 R^2 ~,
\end{equation}
where $C$ is a constant. The curvature is positive when $E < 0$ and negative
when $E > 0$. Since the signature of spacetime requires $E \geq 0$, we follow
only the $+$ case. Choosing $R(t_0, z) = R$ as the spatial coordinate in this
space, we get:
\begin{equation}\label{5.2}
{\rm d} {s_3}^2 = {\cal S}^2 \left[\frac {{\rm d} R^2} {C^2 R^2} + R^2
\left({\rm d} p^2 + {\rm d} q^2\right)\right].
\end{equation}
Note that only one hypersurface can have the property (\ref{5.1}) (since $E$ is
independent of $t$ while $R$ depends on $t$). Thus, the 3-geometry of a space of
constant $t$ can evolve away from or toward (\ref{5.2}), or through (\ref{5.2}),
but cannot preserve this geometry over a finite time.

\begin{figure}
 \includegraphics[scale=0.45]{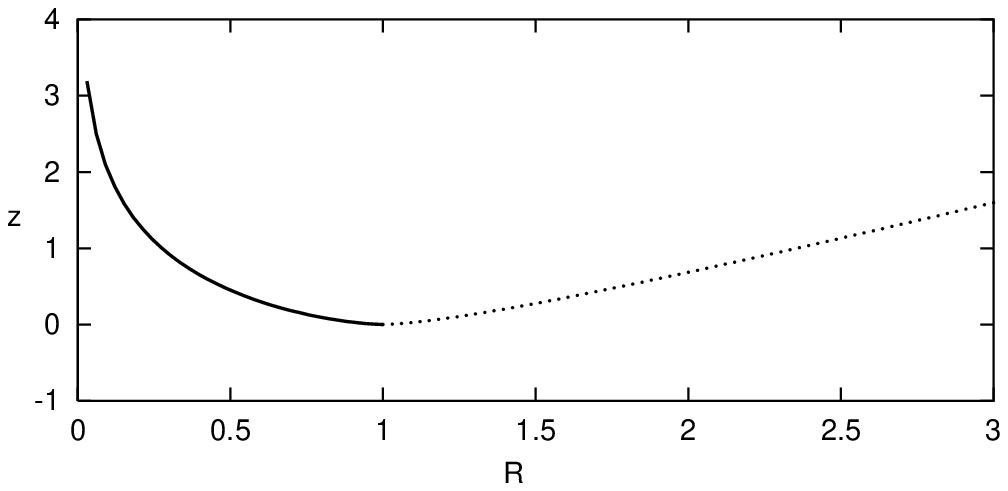}
${}$ \\[-20mm]
 \hspace*{-10mm}
 \includegraphics[scale=0.6]{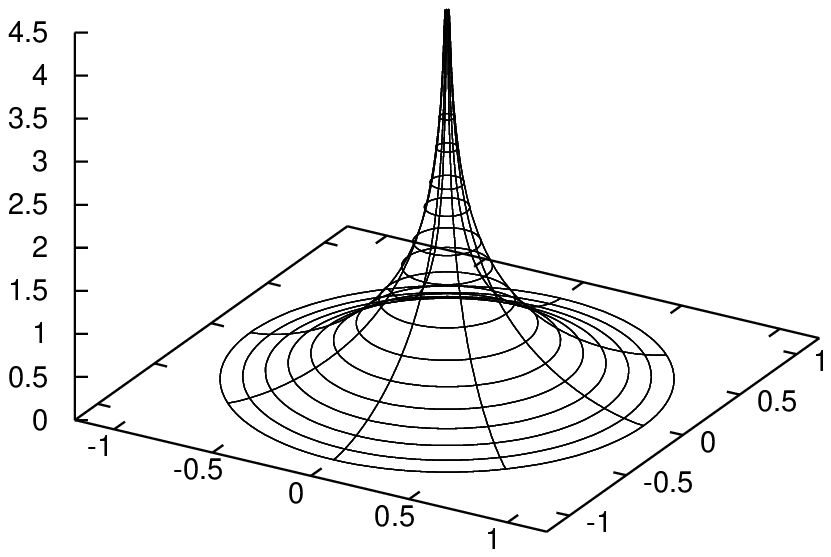}
 \caption{
 \label{embed1}
 \footnotesize
{\bf Top panel:} The function $Z_0(R)$ (from $0$ to $1$, solid line) and the
function $Z_1(R)$ (from $1$ to $3$, dotted line), given by eqs. (\ref{5.4}) and
(\ref{5.6}), respectively. In the graph we chose the $-$ sign for $Z_0(R)$ and
the $+$ sign for $Z_1(R)$. The graph is the cross-section of the $(R,
p)$-surface in the spaces (\ref{5.3}) (left part of the curve) and (\ref{5.5})
(right part of the curve). {\bf Bottom panel:} The $(R, p)$ (or $(R, q)$)
surface obtained by rotating the graph from the top panel around the $R = 0$
axis. The lower end of the funnel is where the embedding in the Euclidean space
breaks down, i.e. where the solid line meets the dotted line in the top panel.
Upwards, the funnel goes infinitely far and becomes infinitely thin. }
 \end{figure}

The surface of constant $q$ in (\ref{5.2}) has the metric ${\rm d} {s_2}^2 =
{\rm d} R^2 / \left(C^2 R^2\right) + R^2 {\rm d} p^2$. To visualise it, we embed
it now in a 3-dimensional Euclidean space with the metric
\begin{equation}\label{5.3}
{\rm d} {s_3}^2 = {\rm d} Z^2 + {\rm d} R^2 + R^2 {\rm d} p^2.
\end{equation}
Our  ${\rm d} {s_2}^2$ is the metric of the surface $Z = Z_0(R)$, where
${Z_{0,R}}^2 + 1 = 1 / (CR)^2$, thus
\begin{eqnarray}\label{5.4}
Z_0 &=& \pm {\displaystyle {\int \frac {\sqrt{1 - C^2 R^2}} {CR}}} {\rm d} R
\\
&=& \pm \frac 1 C \left[\ln {\displaystyle {\left(\frac {CR} {1 + \sqrt{1 - C^2
R^2}}\right)}} + \sqrt{1 - C^2 R^2}\right]. \nonumber
\end{eqnarray}
This embedding is possible only in the range $R \leq 1 / C$. The $R > 1 / C$
part of the surface can be embedded in a flat space of signature $(- +
+)$.\footnote{A similar phenomenon is known from the maximally extended Reissner
-- Nordstr\"{o}m spacetime, when the region inside the interior horizon is
depicted, see Ref. \cite{PlKr2006}.} The metric is then
\begin{equation}\label{5.5}
{\rm d} {s_3}^2 = - {\rm d} {Z_1}^2 + {\rm d} R^2 + R^2 {\rm d} p^2,
\end{equation}
and the embedding equation is
\begin{eqnarray} \label{5.6}
Z_1 &=& \pm \int {\displaystyle {\frac {\sqrt{C^2 R^2 - 1}} {CR}}} {\rm d} R =
\pm \frac 1 C \left[\sqrt{C^2 R^2 - 1} \right. \nonumber \\
&-& 2 \left. \arctan \left(CR + \sqrt{C^2 R^2 - 1}\right) + \frac {\pi} 2\right]
\end{eqnarray}
(the constant of integration was chosen so that $Z_0(1/C) = Z_1(1/C)$). The
functions $Z(R)$ and $Z_1(R)$ are shown in Fig. \ref{embed1}. Note that in both
embeddings, (\ref{5.3}) and (\ref{5.5}), $p$ appears as the polar angle in the
plane $(R, p)$. If $p$ is to be interpreted as actually being a polar angle,
with the period $2 \pi$, then all points with the coordinates $(t, z, p + 2\pi
n, q)$, where $n$ is any integer, should be identical with the point of
coordinates $(t, z, p, q)$. Since $p \to (p +$ constant) are symmetry
transformations of the spacetime (3.1), there is no problem with such an
identification. Thus we should imagine the $(R, p)$ surface as being created by
rotating the curve from Fig. \ref{embed1} around the $Z$ axis.

However, the same picture would be obtained for an $(R, q)$ surface in
(\ref{5.2}), given by $p =$ const. We would find that in that surface, $q$ is
the angular coordinate of the polar coordinates $(R, q)$, and points of
coordinates $(t, z, p, q + 2\pi m)$ can be identified with the point of
coordinates $(t, z, p, q)$. We are thus led to conclude that $(p, q)$ are both
angular coordinates with the period $2 \pi$, and that the points of coordinates
$(p, q)$ have to be identified with the points of coordinates $(p + 2n \pi, q +
2 m \pi)$, where $n$ and $m$ are arbitrary integers. The tentative conclusion is
that the $(p, q)$-surface is a flat torus.

The conclusion is tentative in the sense that, while we identify the set $p =
p_0$ with the set $p = p_0 + 2\pi$, we are still free to carry out the symmetry
transformations within the set $p = p_0$. Thus, the identification can possibly
be done with a twist, that will turn a square into a M\"{o}bius strip, or with a
two-way twist, that will turn it into a projective plane \cite{Bors1964}. We
will use the term ``toroidal topology'' that will be meant to include an
ordinary torus, and also the identifications with twists.

The conclusions drawn from an embedding can be misleading. As an example,
consider the hyperbolically symmetric counterpart of (\ref{3.2}):
\begin{equation}\label{5.7}
{\rm d} s^2 = {\rm d} t^2 - \frac {{R,_z}^2} {2E(z) - 1} {\rm d} z^2 - {R}^2
\left({\rm d} \vartheta^2 + \sinh^2 \vartheta {\rm d} \varphi^2\right).
\end{equation}
The surface of constant $t$ and constant $\varphi$ has the metric ${\rm d}
{s_2}^2 = {R,_z}^2 {\rm d} z^2 /(2E (z) - 1) - {R}^2 {\rm d} \vartheta^2$, and
embedding it in a Euclidean space we would conclude that $\vartheta$ is the
polar coordinate with the period $2\pi$. However, in this case $\vartheta \to
(\vartheta +$ constant) are not symmetry transformations of the spacetime (or of
a space of constant $t$), and so identifications of points with different values
of $\vartheta$ are not permitted. Thus, the embedding in this case is not a
one-to-one mapping. Consequently, the toroidal interpretation of the plane
symmetric case must be treated as one possibility, and not as a definitive
conclusion.

Note from (\ref{5.2}) that the length of any segment of a curve given by $p =$
const and $q =$ const that goes into the point $R = 0$ is infinite. The length
of such a ``radial'' line between the values $R_1$ and $R_2$ is
 $$
\ell_{12} = \left|\int_{R_1}^{R_2} \frac {{\rm d} R} {CR}\right| = \left|\frac 1
C\ \ln \left(\frac {R_2} {R_1}\right)\right| \llim{R_2 \to 0} \infty
 $$
Thus, Fig. \ref{embed1} correctly suggests that a surface of constant $p$ or
constant $q$ in the space (\ref{5.2}) has the shape of an infinite funnel, and
the point with coordinate $R = 0$ is not accessible (does not in fact belong to
this surface). This conclusion is consistent with the observation made in Ref.
\cite{HeKr2008} that in the planar Szekeres metric ``there is no real origin,
but $R$, $M$ and $E$ can asymptotically approach zero''\footnote{In this quote,
notation has been adapted to that used here.}

\begin{figure}
 \begin{center}
 ${}$ \\[1.5cm]
 \hspace*{-1cm}
  \includegraphics[scale=0.58]{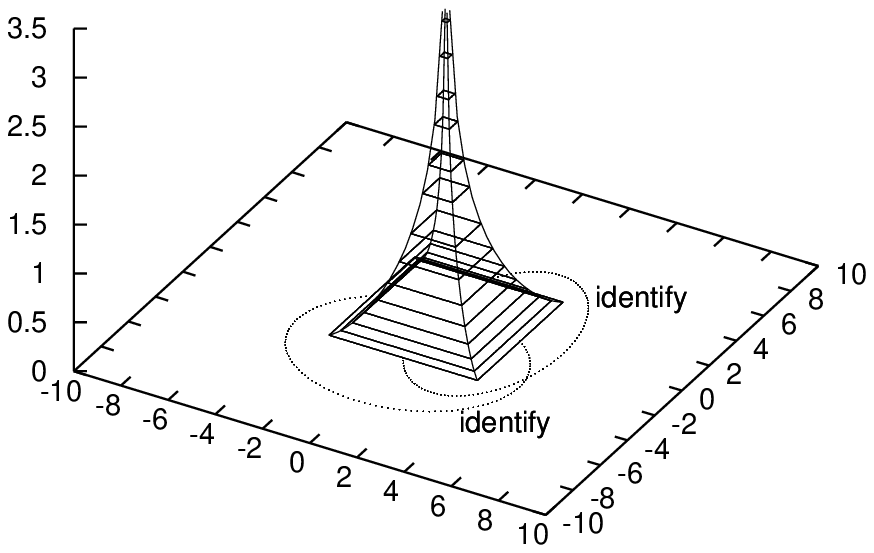}
 \caption{
 \label{fulltorus}
 \footnotesize
A sketch of the 3-space of the plane symmetric to\-ro\-idal model (a faithful
picture cannot be drawn because the 2-dimensional flat torus cannot be embedded
in a Euclidean space, and the 3-space of a planar Szekeres model cannot be made
flat). Each square section of the funnel represents a 2-dimensional flat torus,
so its front edge coincides with the back edge, and the left edge coincides with
the right edge. Each torus is an orbit of the symmetry group of the model. Each
smaller torus is contained within all the larger ones, but the inclusion
relation cannot be depicted in the Euclidean space of this graph. Also, in the
curved 3-space, the 2-tori surround the asymptotic origin, which is the tip of
the funnel, situated infinitely high above the plane shown.}
 \end{center}
 \end{figure}

It can be concluded from (\ref{3.2}) that the $(p, q)$ surface should have a
toroidal topology with any form of $E$, as will now be shown. The 3-metric of a
$t = t_0$ space is:
\begin{equation}\label{5.8}
{\rm d} {s_3}^2 = \frac {{\rm d} {R_0}^2} {2E} + {R_0}^2 \left({\rm d} p^2 +
{\rm d} q^2\right),
\end{equation}
where $R_0(z) \df R(t_0, z)$. We can now embed a surface of constant $p$ or a
surface of constant $q$ in a 3-dimensional flat space by the same method that we
used for (\ref{5.2}), only the equation of embedding will not be explicit:
\begin{equation}\label{5.9}
\pm {Z,_R}^2 + 1 = 1/(2E) > 0,
\end{equation}
the upper sign being for embedding in the Euclidean space, the lower sign for
the embedding in the pseudoeuclidean space. In each case the coordinates $p$ and
$q$ turn out to be the azimuthal coordinates. As argued in Ref. \cite{HeKr2008},
if a nonsingular origin (where $R = E = 0$) is to exist, then it will be
infinitely far from every point of the $t =$ const space. This implies the
infinite funnel geometry of Fig. \ref{embed1}. A sketch of such a space is shown
in Fig. \ref{fulltorus}.

The toroidal geometry and topology of the $(p, q)$ surfaces neatly explains the
pattern of expansion. The circumference of the torus along the $p$- or
$q$-direction is $2 \pi R$ in the coordinates of (\ref{5.8}). Thus, as $R$
increases with time, the circumference of the torus increases in proportion to
$R$, which causes transversal expansion.

The toroidal topology also solves the problem of where the mass generating the
gravitational field resides. As observed in Ref. \cite{HeKr2008}, the regularity
conditions at an origin $z = z_c$ are independent of $\varepsilon$ (in some
cases they cannot be fulfilled, for example with $\varepsilon < 0$). Thus,
$E/M^{2/3}$ must tend to a nonzero constant as $z \to z_c$. Knowing this, let us
calculate the amount of rest mass in an arbitrary volume ${\cal V}$, from
(\ref{3.3}) and (\ref{3.2}). That amount is ${\cal M} \df \int_{\cal V} \rho
\sqrt{g_3} {\rm d}_3x$, where $g_3$ is the determinant of the 3-metric of a $t
=$ const subspace of (\ref{3.2}). Thus
\begin{equation}\label{5.10}
{\cal M} = \frac {c^2} {4 \pi G} \int_{\cal V} \frac {M,_z} {\sqrt{2E}} {\rm d}
p {\rm d} q  {\rm d} z \equiv \frac {c^2} {4 \pi G} \int_{\cal V} \frac 1
{\sqrt{2E}} {\rm d} p {\rm d} q  {\rm d} M.
\end{equation}
With $0 < E \propto M^{2/3}$ in the vicinity of $M = 0$, the integral with
respect to $M$ is finite. With a toroidal topology, the ranges of $p$ and $q$
are finite, so the integrals over $p$ and $q$ also give final values. Thus, the
total amount of mass in each space $t =$ const is finite.

The relation ${\cal M},_z = M,_z/\sqrt{2E}$ that follows from (\ref{5.10}) is
similar to ${\cal M},_r = M,_r/\sqrt{1 + 2E}$, which held in the spherically
symmetric and quasi-spherical models. By analogy, we conclude that in a plane
symmetric spacetime the factor $1/\sqrt{2E}$ measures the relativistic mass
defect/excess, i.e. the discrepancy between the active gravitational mass $M$
and the sum of rest masses ${\cal M}$.

\section{No apparent horizons in the quasi-plane and quasi-hyperbolic models}
\label{noAH}

\setcounter{equation}{0}

An apparent horizon is the envelope of the region of trapped surfaces. A (past
or future) trapped surface is such, on which both the ingoing and outgoing
(past- or future-directed respectively) families of null geodesics converge. A
future AH always forms in spherically symmetric or quasi-spherical-Szekeres
collapse before the Big Crunch singularity is achieved, a past AH always exists
after the Big Bang singularity.

It turns out that the AH-s do not exist in the quasi-plane and quasi-hyperbolic
Szekeres models, and, consequently, neither do they exist in the plane- and
hyperbolically symmetric dust models. Actually, a stronger result holds: these
spacetimes remain trapped for all the time. This follows by the method invented
by Szekeres \cite{Szek1975b}, which applies here almost unchanged -- only the
final conclusion is radically different in consequence of the different sign of
$\varepsilon$. To avoid getting into complicated details, we begin by using the
general form (\ref{2.1}) of the metric. Suppose a trapped surface exists, and
call it $\Sigma$.

We assume $\Sigma$ to be one of the orbits of the quasi-symmetry, i.e. to have
its equation of the form $\{t = {\rm constant}, z = {\rm constant}\}$. It will
be explained later (see after (\ref{6.9})) why it is sufficient to consider such
surfaces to prove the conclusion. The traditional definition of a trapped
surface requires that it be compact. With the toroidal topology in the planar
model, our $\Sigma$ will be compact indeed. With the infinite topology, and in
the quasi-hyperbolic model, $\Sigma$ will be infinite. In view of the final
result of our consideration, this fact will turn out to be unimportant. We
choose these infinite surfaces because of their simple geometry.

Consider any family of null geodesics intersecting $\Sigma$ orthogonally, and
let the tangent vector field of those geodesics be $k^{\mu}$. Let $(t, z, x, y)
= (x^0, x^1, x^2, x^3)$. We have then
\begin{equation}\label{6.1}
k_{\mu} k^{\mu} = 0, \qquad k^{\nu} {k^{\mu}};_{\nu} = 0 \qquad {\rm everywhere}
\end{equation}
because $k^{\mu}$ is tangent to null geodesics, and
\begin{equation}\label{6.2}
k^2 = k^3 = 0, \qquad \left(k^0\right)^2 - {\rm e}^{2 \alpha} \left(k^1\right)^2
= 0 \qquad {\rm on\ } \Sigma
\end{equation}
because $k^{\mu}$ is assumed orthogonal to $\Sigma$, so at the points of
$\Sigma$ it must be spanned on the vector fields normal to $\Sigma$, which are
$(1, 0, 0, 0)$ and $(0, {\rm e}^{- \alpha}, 0, 0)$. The affine parameter along
each null geodesic may be chosen so that
\begin{equation}\label{6.3}
k^0 = {\rm e}^{\alpha}, \qquad k^1 = e = \pm 1 \qquad {\rm on\ } \Sigma,
\end{equation}
where we will call the geodesics with $e = -1$ ``ingoing'', and those with $e =
+1$ ``outgoing''.\footnote{When the surface of constant $t$ and $z$ is infinite,
it cannot be closed, therefore the labelling ``ingoing'' and ``outgoing'' is
only conventional.} A surface $\Sigma$ is trapped when the expansion
${k^{\mu}};_{\mu}$ calculated on $\Sigma$ is negative for both families. We have
on $\Sigma$, using (\ref{6.2}):
\begin{equation}\label{6.4}
{k^{\mu}};_{\mu} = {k^0},_t + {k^1},_z + {\rm e}^{\alpha} \left(\alpha,_t + 2
\beta,_t\right) + e \left(\alpha,_z + 2 \beta,_z\right).
\end{equation}
In order to simplify this, we now differentiate the first of (\ref{6.1}) by $t$,
and write out the second of (\ref{6.1}) for $\mu = 1$, in both cases taking the
result on $\Sigma$, i.e. making use of the simplifications given in (\ref{6.2}):
\begin{eqnarray}\label{6.5}
&& {k^0},_t - e {\rm e}^{\alpha} {k^1},_t - {\rm e}^{\alpha} \alpha,_t = 0,
\nonumber \\
&& {\rm e}^{\alpha} {k^1},_t + e \left({k^1},_z + 2 {\rm e}^{\alpha}
\alpha,_t\right) + \alpha,_z = 0.
\end{eqnarray}
Eliminating ${k^1},_t$ from (\ref{6.5}), and using the result to substitute for
${k^0},_t + {k^1},_z$ in (\ref{6.4}) we get
\begin{equation}\label{6.6}
{k^{\mu}};_{\mu} = 2 \left({\rm e}^{\alpha} \beta,_t + e \beta,_z\right).
\end{equation}
Using now the expressions for ${\rm e}^{\alpha}$ and ${\rm e}^{\beta}$ in the
notation of (\ref{2.7}), i.e.
\begin{equation}\label{6.7}
{\rm e}^{\alpha} = \frac {R,_z - R {\cal E},_z/{\cal E}} {\sqrt{\varepsilon +
2E(z)}}, \qquad {\rm e}^{\beta} = \frac R {\cal E}
\end{equation}
we get in (\ref{6.6})
\begin{equation}\label{6.8}
{k^{\mu}};_{\mu} = 2 \left(\frac {R,_z} R - \frac {{\cal E},_z} {\cal E}\right)
\left(\frac {R,_t} {\sqrt{\varepsilon + 2E}} + e\right).
\end{equation}
The first factor changes sign only at shell crossings (see Ref.
\cite{HeKr2002}), so we take it to be positive. Consider collapse, $R,_t < 0$.
For the ingoing family, $e = -1$, we have ${k^{\mu}};_{\mu} < 0$, without
further conditions. For the outgoing family, $e = +1$, ${k^{\mu}};_{\mu}$ will
be negative when $R,_t / \sqrt{\varepsilon + 2E} < -1$, which, with negative
$R,_t$, means that ${R,_t}^2 > \varepsilon + 2E$. Using (\ref{2.8}) with
$\Lambda = 0$ for ${R,_t}^2$, we then obtain
\begin{equation}\label{6.9}
2M/R > \varepsilon.
\end{equation}
With $\varepsilon = 0$ and $\varepsilon = -1$, this is always
fulfilled,\footnote{Note that $M$ must be positive, or else (\ref{3.6}) would
imply that collapse is retarded and expansion accelerated. This would be
gravitational repulsion.} with the only exception of the 'asymptotic origin' in
the planar model, where $M/R = \varepsilon = 0$.

A surface given by  $\{t = {\rm constant}, z = {\rm constant}\}$ passes through
every point of the spacetime. Since each such surface has now been shown to be
trapped at all of its points, this means that all points of the whole spacetime
are trapped.

Thus, the quasi-hyperbolic and quasi-plane model, along with their
hyperbolically- and plane-symmetric limits, are globally future-trapped (when
collapsing), and no apparent horizon exists for them. It follows now easily that
the corresponding expanding models are globally past-trapped.

This is consistent with the fact that the corresponding vacuum solutions have no
event horizons (see eq. (6.22) in Ref. \cite{HeKr2008}) and are globally
nonstatic.

With no apparent horizons, no  black holes may form (more precisely, the whole
Universe is one black hole). This excludes the quasi-plane and quasi-hyperbolic
models from an important area of application of the quasi-spherical and
spherically symmetric dust models.

\section{The toroidal plane symmetric model in the Szekeres
coordinates}\label{torpsSzco}

\setcounter{equation}{0}

\begin{figure}
 \begin{center}
 \hspace*{-4mm}
 \includegraphics[scale = 0.5]{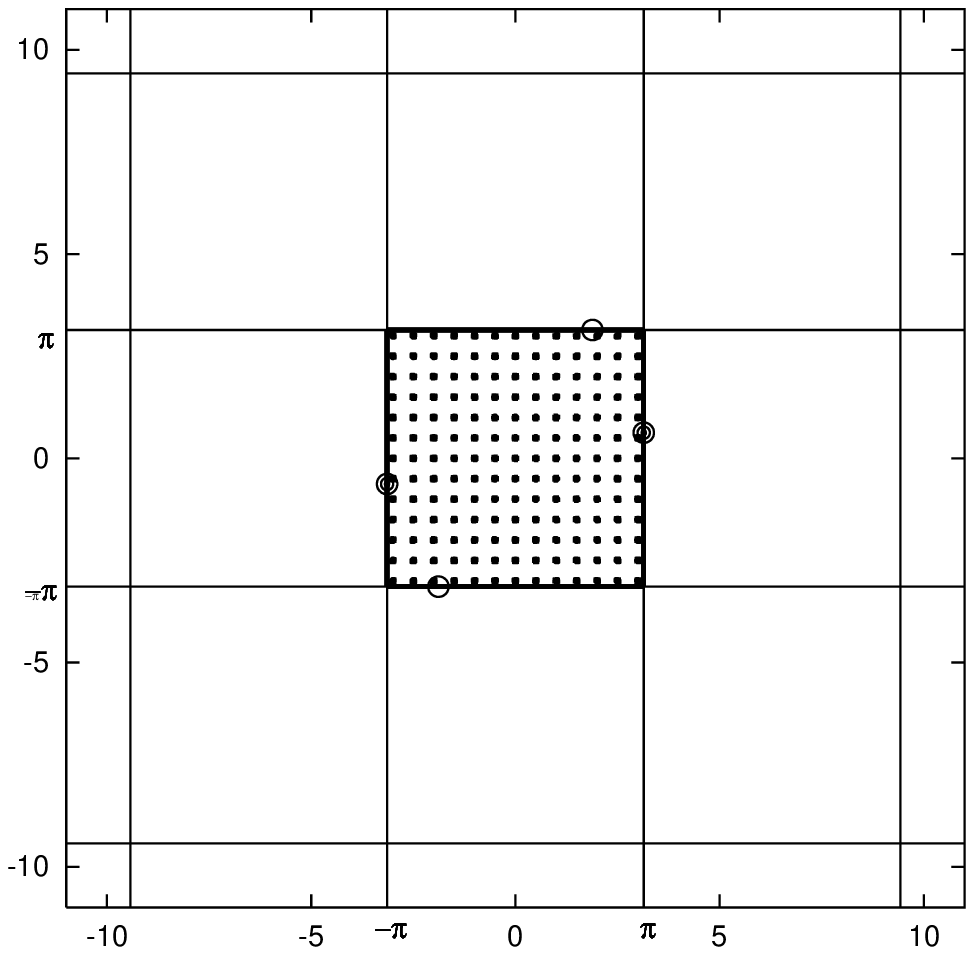}
 \caption{
 \label{torusmap}
 \footnotesize
A map of the elementary torus in the $(p, q)$ coordinates (the central dotted
square). Its left edge coincides in space with the right edge, the lower edge
coincides with the upper edge. Sometimes it is convenient to consider the torus
as a subset of the $\mathbb{R}^2$ plane, in which case the plane should be
imagined as covered with an infinite number of copies of the elementary square.
The identifications may be done with a twist, i.e. with reflections in $p = 0$
and $q = 0$, so that a projective plane results instead of a torus. The open
circles and the full circles show the pairs of points to be identified in the
latter situation.
 }
 \end{center}
 \end{figure}

The metric of a torus given by constant $t$ and $z$ in (\ref{3.2}) is ${\rm d}
{s_2}^2 = R^2 \left({\rm d} p^2 + {\rm d} q^2\right)$, where $2 \pi R$ is the
circumference of the torus in the $p$-direction and in the $q$-direction. In the
following, we will consider the torus with $R = 1$, and we will call it the
'elementary square' or 'elementary torus'. It will be more convenient to assume
that, in the coordinates of (\ref{3.2}), the elementary torus is the square
$\{p, q\} \in [- \pi, \pi] \times [- \pi, \pi]$, shown in Fig. \ref{torusmap},
rather than $\{p, q\} \in [0, 2 \pi] \times [0, 2 \pi]$. The
$\mathbb{R}^2$-space of the $(p, q)$ coordinates can be imagined as filled with
infinitely many copies of this square.

As observed in Ref. \cite{HeKr2008}, the function $S$ in the quasi-plane model
can be absorbed into the other functions by the redefinition
\begin{equation}\label{7.1}
(R, E, M) = (\widetilde{R}/S, \widetilde{E}/S^2, \widetilde{M}/S^3),
\end{equation}
so we can assume $S = 1$ with no loss of generality. We do this in the
following.

The coordinates of points to be identified are related in a more complicated way
in the Szekeres coordinates of (\ref{2.1}), in which the plane symmetric model
is given by:
\begin{equation}\label{7.2}
{\rm d} s^2 = {\rm d} t^2 - \frac {{R,_z}^2} {2E(z)}  {\rm d} z^2- {R}^2 \frac
{4 \left({\rm d} x^2 + {\rm d} y^2\right)} {\left[\left(x - P\right)^2 + \left(y
- Q\right)^2\right]^2},
\end{equation}
with $P$ and $Q$ being arbitrary constants. A line $q = q_0$ corresponds, in the
$(x, y)$-coordinates, to
\begin{equation}\label{7.3}
(x - P)^2 + (y - Q - 1/q_0)^2 = 1/{q_0}^2,
\end{equation}
which is in general a circle of radius $1/q_0$ and the center at $(x, y) = (P, Q
+ 1/q_0)$. In the special case $q_0 = 0$ the image becomes the straight line $y
= Q$. Consequently, the lines $q = \pm \pi$ in the $(p, q)$-coordinates go over
into the circles
\begin{equation}\label{7.4}
(x - P)^2 + (y - Q \mp 1/\pi)^2 = 1/\pi^2,
\end{equation}
while the lines $p = \pm \pi$ go over into the circles
\begin{equation}\label{7.5}
(x - P \mp 1/\pi)^2 + (y - Q)^2 = 1/\pi^2.
\end{equation}
The image of the central point $(p, q) = (0, 0)$ is the infinity of the $(x,
y)$-plane. Conversely, the point $(x, y) = (P, Q)$ is the image of the infinity
of the $(p, q)$-coordinates.

\begin{figure}
 \begin{center}
 \hspace*{-5mm}
 \includegraphics[scale = 0.5]{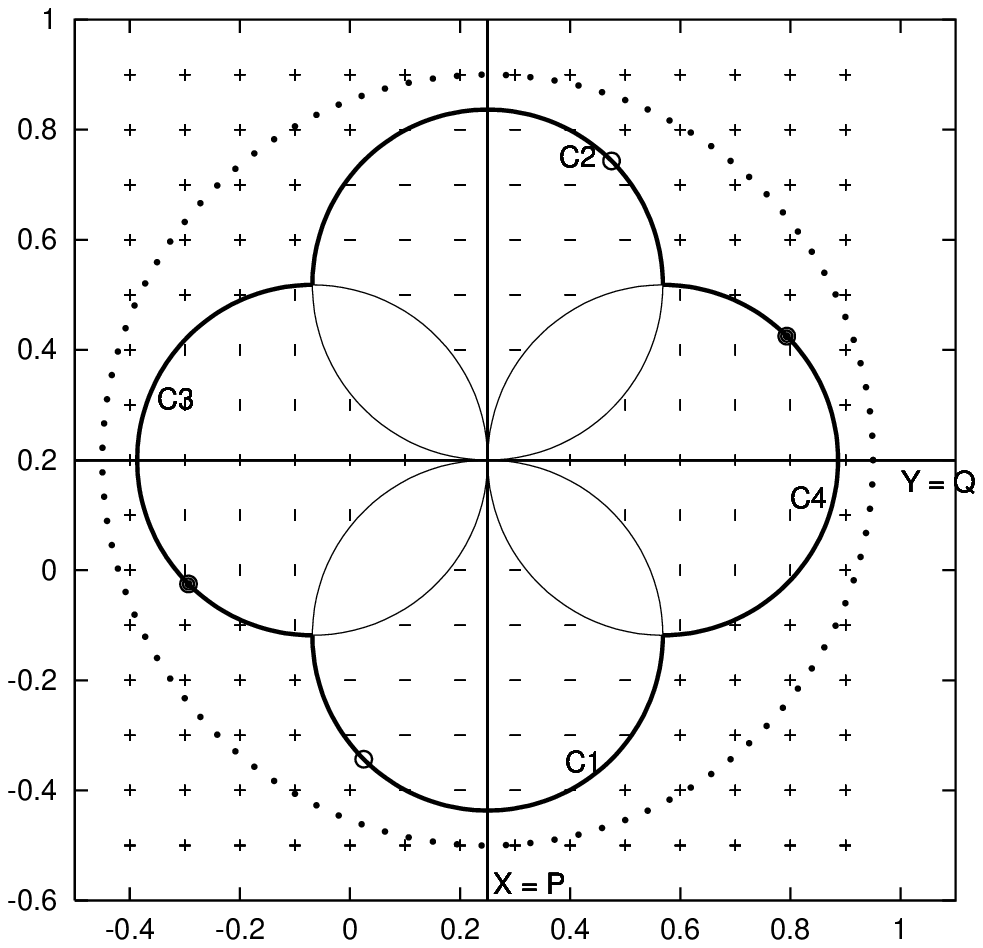}
 \caption{
 \label{torusimage}
 \footnotesize
The image of the torus $\{p, q\} \in [- \pi, \pi] \times [- \pi, \pi]$ in the
Szekeres coordinates. The values of $P$ and $Q$ were chosen arbitrarily, but
other elements of the figure are drawn to scale. The small empty circles and the
small solid circles mark pairs of points to be identified when we consider a
projective plane instead of a torus. More explanation in the text.
 }
 \end{center}
 \end{figure}

Moreover, from (\ref{7.3}) follows that the image of the area $\{q^2 <
{q_0}^2\}$ (an infinite strip of the $(p, q)$ plane contained between $q = -
q_0$ and $q = q_0 > 0$) is the area \textit{outside} the circles $(x - P)^2 + (y
- Q \mp 1/q_0)^2 = 1 / {q_0}^2$. Similarly, the image of the area $\{p^2 <
{p_0}^2\}$ is the area outside the circles $(x - P\mp 1/p_0)^2 + (y - Q )^2 = 1
/ {p_0}^2$. Consequently, the image of the elementary torus in the $(x, y)$
coordinates will be the infinite subset of the $\mathbb{R}^2$ plane lying
outside all four circles, see Fig. \ref{torusimage}.

This is the explanation to Fig. \ref{torusimage}. The image of the line $p = 0$
in the $(x, y)$-coordinates is the vertical line $x = P$ in the figure, with
$(x, y) = (P, Q)$ being the image of infinity of the $(p, q)$ coordinates.
Similarly, the image of $q = 0$ is the horizontal line $y = Q$. In the $(p,
q)$-coordinates of (\ref{3.2}), the torus is the area encircled by the straight
lines $p = - \pi$, $p = \pi$, $q = - \pi$ and $q = \pi$. The image of the line
$q = - \pi$ is the circle C1, of radius $1/\pi$. The image of the line $q = \pi$
is the circle C2, of the same radius. The image of the torus must be contained
outside these two circles -- in the area covered with vertical strokes. Then,
the image of $p = - \pi$ is the circle C3, and the image of $p = \pi$ is the
circle C4, both of the same radius $1/\pi$. Consequently, the image of the torus
must be contained outside these circles -- in the area covered by horizontal
strokes. Thus, the image of the whole torus is the common subset of these two
areas -- the area in the figure outside the thick line and covered with crosses.
The area inside the thick closed curve contains an infinite number of images of
copies of the elementary torus.

Each circle in the figure with the center at $(x, y) = (P, Q)$ and with radius
$a > 2 / \pi$ is an image of a circle of radius $2/a < \pi$ centered at $(p, q)
= (0, 0)$ that lies all within a single copy of the elementary torus (one such
circle is shown in the figure with a dotted line). In particular, this applies
to a circle of radius $1$.

\section{A nonsymmetric planar model}\label{torplnonsym}

\setcounter{equation}{0}

The intention of the foregoing considerations was to prepare the ground for
carrying out similar identifications in the full nonsymmetric quasi-plane model.
Unfortunately, no such identifications were proven possible. The claim made in
Ref. \cite{Kras2008} turned out to be erroneous, in consequence of a
computational error. That error invalidates the statements made in Ref.
\cite{Kras2008} in sections VIII, IX and XI, which are therefore not included in
the present text. Section VIII contained the incorrect identification, Section
IX -- the prescription to avoid shell crossings with a toroidal topology, and
Section XI -- a prescription for constructing a two-sided (orientable) compact
surface out of four copies of a projective plane, which is nonorientable. The
surface thus constructed was supposed to contain the active gravitational mass
in a similar way to the one described in Sec. V for tori.

It has not been proved that nontrivial topologies for the surfaces of constant
$(t, z)$ are possible or impossible in the metric (\ref{2.7}) with $\varepsilon
= 0$, so the problem is still open.

\section{Formation of structures in the planar model}\label{struform}

\setcounter{equation}{0}

This was Section X in Ref. \cite{Kras2008}.

As shown for the Lema\^{\i}tre -- Tolman models (see Refs. \cite{Silk1977} and
\cite{PlKr2006}, Sec. 18.19), in the ever-expanding case $E > 0$ an increasing
density perturbation, $\rho,_z / \rho$, freezes asymptotically into the
background -- i.e. it tends to a finite value determined uniquely by the initial
conditions. Consequently, it is impossible in these models to describe the
formation of condensations that collapse to a very high density, such as a
galaxy with a central black hole. Since the evolution of the quasi-plane and
quasi-hyperbolic models is described by the same equations, they will suffer
from the same problem. (And we have already found in Sec. \ref{noAH} that these
models cannot describe black holes.)

Thus, these models can be used for considering the formation of
moderate-amplitude condensations and voids.

\section{Interpretation of $M(z)$ for the quasi-plane model with infinite
spaces}\label{infmassint}

\setcounter{equation}{0}

With the toroidal interpretation in the plane symmetric limit, the proof that
$M(z)$ is a measure of the active gravitational mass was rather simple. The same
may be shown also when the quasi-plane model is interpreted as infinite in
extent, but in a more complicated way. We do show it in this section -- however,
this is only for mathematical completeness. As demonstrated in Ref.
\cite{HeKr2008}, with the infinite spaces the quasi-plane Szekeres solutions
have irremovable shell crossings, and so are in fact not acceptable as
cosmological models.

This section was Sec. XII in Ref. \cite{Kras2008}.

Recall that with $\varepsilon = 0$ we are free to rename the functions $R$, $E$
and $M$ as in (\ref{7.1}), and the result will be the same as if $S \equiv 1$.
Thus, we assume $S \equiv 1$ throughout this section.

For the beginning let us consider the plane symmetric subcase of the
$\varepsilon = 0$ model, which has $P,_z = Q,_z = 0$. Let us choose a circle of
radius 1 centred at $(x, y) = (P, Q)$ (both $P$ and $Q$ being now constant) in
every surface of constant $t$ and $z$. Let ${\rm d}_2 xy$ be the surface element
in the $(x, y)$ plane, and let $U$ be the outside of the unit circle. This
region has finite surface area. Then, introducing $(u, \varphi)$ by
\begin{equation}\label{10.1}
x = P + u \cos \varphi, \qquad y = Q + u \sin \varphi,
\end{equation}
we get
\begin{eqnarray}
\int_U {\rm d}_2 xy\ \frac 1 {{\cal E}^2} = \int_0^{2 \pi} {\rm d} \varphi
\int_1^{\infty} \frac 4 {u^3} {\rm d} u = 4 \pi, \label{10.2}
\end{eqnarray}
in every $(t = {\rm const}, z = {\rm const})$ surface. Now let ${\cal V}$ be a
3-dimensional set in a $t =$ const space, extending from $z = z_c$ to a running
value of $z$, whose every section of constant $z$ is $U$ -- the outside of the
unit circle $(x - P)^2 + (y - Q)^2 = 1$. Calculating ${\cal M} = \int_{\cal V}
\rho \sqrt{\left|g_3\right|} {\rm d}_3 x$ with $\varepsilon = 0$, $S = 1$ and
$P, Q$ constant we get ${\cal E},_z = 0$ and
\begin{eqnarray} \label{10.3}
&& {\cal M} = \frac 1 {4 \pi} \int_U {\rm d}_2 xy \int_{z_c}^z {\rm d} u \frac
{M,_u} {\sqrt{2E} {\cal E}^2} = \int_{z_c}^z \frac {M,_u} {\sqrt{2E}}(x) {\rm d}
u, \nonumber \\
&&
\end{eqnarray}
Thus, in this case, $M$ behaves as the active gravitational mass contained
outside a tube of coordinate height $(z - z_c)$ which has radius equal to 1 at
every $z$ value, while $1 / \sqrt{2E}$ plays the role of the mass defect/excess
factor. We recall that the $(x, y)$ coordinates of (\ref{2.7}), in the plane
symmetric case $\varepsilon = 0 = {\cal E},_z, S = 1$ are related by the
inversion (\ref{3.1}) to the Cartesian coordinates $(p, q)$ in a plane, so the
outside of the tube in the $(x, y)$ coordinates is in reality the inside of the
same tube in the Cartesian coordinates. Thus, physically, $M$ is the active mass
within a tube ${\cal V}_2$.

We calculated the integrals in (\ref{10.2}) and (\ref{10.3}) around the central
point $(x, y) = (P, Q)$. However, with plane symmetry, the origin of the
Cartesian coordinates can be transferred to any other point by a symmetry
transformation.\footnote{The circle of unit radius in the Cartesian coordinates,
when moved to another point of the $(p, q)$ plane, will not have a unit
coordinate radius in the $(x, y)$ coordinates, and the image of the center of
the circle will not be the center of the image circle. However, the surface area
of the circle and the invariant distances between points are not changed.}

Let us consider the transformation (\ref{3.1}), after which the metric becomes
(\ref{3.2}), which is formally (\ref{2.7}) with ${\cal E} = 1$ and $(x, y)$
renamed to $(p, q)$. In this form, the transformation
\begin{equation}
p = p' + A_p, \qquad q = q' + A_q \label{10.4}
\end{equation}
(with $A_p$ and $A_q$ being arbitrary constants) is an isometry of (\ref{3.2}).
Thus, the transformation (\ref{10.4}) does not change either the metric
(\ref{3.2}) or the value of the integral (\ref{10.2}), which, in the variables
$(p, q)$, becomes simply $4 \int_{S_1} {\rm d} p {\rm d} q = 4\pi$,
independently of where the centre of the circle $S_1$ is located.

Now let us consider the planar metric $\varepsilon = 0$ that is not plane
symmetric, i.e. with $P,_z$ and $Q,_z$ not vanishing simultaneously. Let $U(z)$
be the outside of a unit circle in an $z =$ const surface, with the centre at
$(x, y) = (P(z), Q(z))$. Within each single such surface, applying the
transformation of variables (\ref{10.1}), we get
\begin{eqnarray} \label{10.5}
&& \int_U {\rm d}_2 xy\ \frac 1 {{\cal E}^2} = \int_0^{2 \pi} {\rm d} \varphi
\int_1^{\infty} \frac 4 {u^3} {\rm d} u = 4 \pi, \nonumber \\
&& \int_U {\rm d}_2 xy\ \frac {{\cal E},_z} {{\cal E}^3} \\
&& = \int_0^{2 \pi} {\rm d} \varphi \int_1^{\infty} \frac {- 4 u \cos \varphi
P,_z - 4 u \sin \varphi Q,_z} {u^5}\ {\rm d} u = 0. \nonumber
\end{eqnarray}
These integrals do not depend on $P,_z$ or $Q,_z$, but the centres of the
circles no longer have the same $(x, y)$ coordinates at each $z$. Thus, to use
(\ref{10.5}) in an analogue of (\ref{10.3}), we have to take a volume ${\cal V}$
which is a wiggly tube: its every cross-section with a constant $z$ surface is a
unit circle, but the centres of the circles do not lie on a line orthogonal to
the $z =$ const surfaces. Instead, they lie on the curve in the $t =$ const
space given by the parametric equations $x = P(z), y = Q(z)$. Because of the
second of (\ref{10.5}), (\ref{10.3}) still follows for this single tube.

The whole 3-space $t =$ const is now no longer homogeneous with respect to the
group of plane symmetries. However, each single $z =$ const surface in that
space is homogeneous. In particular, the surface containing the base of the
tube, $z = z_0$, is homogeneous. Thus, we can apply the inversion (\ref{3.1})
with $P = P(z_0), Q = Q(z_0)$. The inside and outside of the unit circle in the
$z = z_0$ surface will thereby simply interchange, but the resulting
transformations in other $z =$ const surfaces will be more complicated, and the
wiggly tube will deform substantially. Still, in the inverted coordinates we are
now free to move the centre of the base circle (within the $z = z_0$ surface) to
any other point.

We now carry out this plan. Let us denote:
\begin{eqnarray}\label{10.6}
P(z_0) &\df& P_0, \qquad Q(z_0) \df Q_0, \nonumber \\
V &\df& (P_0 - P)^2 + (Q_0 - Q)^2.
\end{eqnarray}
To the metric (\ref{2.7}) with $\varepsilon = 0$ and $S = 1$ we apply the
inversion adapted to the surface $\{t = {\rm const}, z = z_0\}$:
\begin{equation}\label{10.7}
x = P_0 + \frac p {p^2 + q^2}, \qquad y = Q_0 + \frac q {p^2 + q^2}.
\end{equation}
After this, the 2-metric $R^2 \left({\rm d} x^2 + {\rm d} y^2\right) / {\cal
E}^2$ becomes:
\begin{eqnarray}\label{10.8}
{\rm d} {s_2}^2 &=& \frac 1 {{\widetilde{\cal E}}^2}\ \left({\rm d} x^2 + {\rm
d} y^2\right), \\
2 \widetilde{\cal E} &=& V \left(p^2 + q^2\right) + 2 \left(P_0 - P\right) p + 2
\left(Q_0 - Q\right) q + 1. \nonumber
\end{eqnarray}
In these coordinates, the surface $\{t = {\rm const}, z = z_0\}$ is explicitly
homogeneous, so we are now free to shift the origin of coordinates to any other
point by
\begin{equation}\label{10.9}
p = p' + A_1, \qquad q = q' + A_2,
\end{equation}
with $A_1$ and $A_2$ being constants. After the shift, the metric is still
Szekeres with $\varepsilon = 0$, but with complicated expressions for the new
$P$, $Q$ and $S$.

After the transformations (\ref{10.7}) and (\ref{10.9}) the region $U$ of
integration in (\ref{10.5}) (which was the outside of a tube extending out to
infinity) goes over into a finite region -- the inside of a certain tube whose
edge is the image of the family of unit circles in $(x, y)$. In the integrals
(\ref{10.5}), the two transformations are just changes of integration variables,
so the values of the integrals do not change, and thus (\ref{10.3}) still
applies. This shows that over each point of the surface of constant $t$ and of
$z = z_0$ in the Szekeres $\varepsilon = 0$ metric, we can find a region of
finite volume (a wiggly tube) such that the function $M$ can be interpreted as
the active gravitational mass within that tube.

As an illustration we now consider a special case of the transformation
(\ref{10.9}) with $A_1 \df \lambda$ and $A_2 = 0$. But first we give the
complete transformation that will take us back to the coordinates of (\ref{2.7})
in the surface $z = z_0$.

To the variables $(p', q')$ of (\ref{10.9}) we apply the inversion in a circle
of radius 1 centred at $(p', q') = (0, 0)$, and the shift by $(P_0, Q_0)$ to the
resulting $(x', y')$ coordinates. Calling the final coordinates $(x_2, y_2)$, we
calculate the effect of (\ref{10.9}) on the variables $(x, y)$. The complete
transformation from $(x, y)$ to $(x_2, y_2)$ is
\begin{eqnarray}\label{10.10}
x_2 &=& P_0 + \frac 1 W\ \left(x - P_0 + A_1 {\cal U}\right), \nonumber \\
y_2 &=& Q_0 + \frac 1 W\ \left(y - Q_0 + A_2 {\cal U}\right),
\end{eqnarray}
where
\begin{eqnarray}\label{10.11}
&& \hspace{-5mm} {\cal U} \df (x - P_0)^2 + (y - Q_0)^2, \\
&& \hspace{-5mm} W \df 1 + 2\left[A_1 (x - P_0) + A_2 (y - Q_0)\right] +
\left({A_1}^2 + {A_2}^2\right) {\cal U}. \nonumber
\end{eqnarray}
This set of transformations does not change the metric (\ref{2.7}) in the
hypersurface $z = z_0$, but after the transformations the unit circle ${\cal U}
= 1$ goes over into the circle
\begin{eqnarray}\label{10.12}
&& \left(x_2 - P_0 - A_1/\gamma\right)^2 + \left(y_2 - Q_0 - A_2/\gamma\right)^2
= 1/\gamma^2, \nonumber \\
&&
\end{eqnarray}
where $\gamma \df {A_1}^2 + {A_2}^2 - 1$.

Now we specialize this to the 1-parameter subgroup $A_1 = \lambda$, $A_2 = 0$,
i.e. to the shift along the $p$-direction in (\ref{10.9}). The transformation
(\ref{10.10}) -- (\ref{10.11}) becomes:
\begin{eqnarray}\label{10.13}
&& x_2 = P_0 + \frac{x - P_0 + \lambda \{(x - P_0)^2 + (y - Q_0)^2\}} {W_0},
\nonumber \\
&& y_2 = Q_0 + \frac{y - Q_0} {W_0}, \\
&& W_0 \df \lambda^2 \{(x - P_0)^2 + (y - Q_0)^2\} + 2 \lambda (x - P_0) + 1,
\nonumber
\end{eqnarray}
and its inverse is obtained by replacing $\lambda$ with $(- \lambda)$, i.e.
\begin{eqnarray}\label{10.14}
&& x = P_0 + \frac{x_2 - P_0 - \lambda \{(x_2 - P_0)^2 + (y_2 - Q_0)^2\}}
{\widetilde{W}_0}, \nonumber \\
&& y = Q_0 + \frac{y_2 - Q_0}{\widetilde{W}_0}, \\
&& \hspace{-4mm} \widetilde{W}_0 \df \lambda^2 \{(x_2 - P_0)^2 + (y_2 - Q_0)^2\}
- 2 \lambda (x_2 - P_0) + 1. \nonumber
\end{eqnarray}
The Jacobians of the two transformations are, respectively
\begin{eqnarray}\label{10.15}
\widetilde{J} &=& \left|\pdr{(x_2, y_2)} {(x, y)}\right| = \frac{1}{W_0^2},
\nonumber \\
J &=& \left|\pdr{(x, y)}{(x_2, y_2)}\right| = \frac{1}{\widetilde{W}_0^2}.
 \end{eqnarray}
The following identities are useful in calculaitons:
\begin{eqnarray}\label{10.16}
(x_2 - P_0)^2 + (y_2 - Q_0)^2 &\equiv& \frac{(x - P_0)^2 + (y - Q_0)^2} {W_0},
\nonumber \\
(x - P_0)^2 + (y - Q_0)^2 &\equiv& \frac{(x_2 - P_0)^2 + (y_2 - Q_0)^2}
{\widetilde{W}_0}. \nonumber \\
&&
\end{eqnarray}

The transformation (\ref{10.13}) takes the circle $(x - P)^2 + (y - Q)^2 = u^2$
to a shifted circle with a different radius, namely
\begin{equation}\label{10.17}
(x_2 - A)^2 + (y_2 - B)^2 = \left(u / p_0\right)^2,
\end{equation}
where
 \begin{widetext}
\begin{eqnarray}\label{10.18}
A &\df& \frac {\lambda^2 P_0 \left[(P_0 - P)^2 + (Q_0 - Q)^2 - u^2\right] -
\lambda \left[u^2 + P_0^2 - P^2 - (Q_0 - Q)^2\right] + P} {p_0}, \nonumber \\
B &\df& \frac {\lambda^2 Q_0 \left[(P_0 - P)^2 + (Q_0 - Q)^2 - u^2\right] - 2
\lambda Q_0 (P_0 - P) + Q}{p_0}, \nonumber \\
p_0 &\df& \lambda^2 \left[(P_0 - P)^2 + (Q_0 - Q)^2 - u^2\right] - 2 \lambda
(P_0 - P) + 1
\end{eqnarray}
 \end{widetext}
(in fact, these will be applied with $u = 1$.)

Now using (\ref{10.14}) -- (\ref{10.16}) we find:
\begin{equation}\label{10.19}
(x - P)^2 + (y - Q)^2 = \frac {S_0} {\widetilde{W}_0}\ \left[(x_2 + \alpha)^2 +
(y_2 + \beta)^2\right],
\end{equation}
where
\begin{eqnarray}\label{10.20}
S_0 &\df& 1 - 2 \lambda (P_0 - P) + \lambda^2 \left[(P_0 - P)^2 + (Q_0 -
Q)^2\right], \nonumber \\
\alpha &\df& - P_0 + \frac {P_0 - P - \lambda \left[(P_0 - P)^2 + (Q_0 -
Q)^2\right]} {S_0}, \nonumber \\
\beta &\df& - Q_0 + \frac {Q_0 - Q} {S_0}.
\end{eqnarray}
Note that $S_0$, $\alpha$ and $\beta$ are functions only of $z$, they do not
depend on $x_2$ and $y_2$. We can see that the ${\widetilde{W}_0}^2$ that will
appear in the transformed integral in (\ref{10.5}), $\int_U {\rm d}_2 xy / {\cal
E}^2$, will be canceled by the ${\widetilde{W}_0}^2$ from the Jacobian of the
transformation, and what remains will be
\begin{equation}\label{10.21}
\int \frac{1}{{\cal E}^2} \, {\rm d} x \, {\rm d} y = \frac 4 {{S_0(z)}^2} \int
\frac 1 {\left[(x_2 + \alpha)^2 + (y_2 + \beta)^2\right]^2} {\rm d} x_2 {\rm d}
y_2,
\end{equation}
an integral of exactly the form (\ref{10.5}), except for the additional factor
$1 / {S_0}^2$.

The transformation (\ref{10.13}) affects also the metric. Under (\ref{10.13}),
${\cal E}$ changes as follows:
\begin{equation}\label{10.22}
{\cal E} = (x - P)^2 + (y - Q)^2 = \frac {S_0} {\widetilde{W}_0}\ \left[(x_2 +
\alpha)^2 + (y_2 + \beta)^2\right].
\end{equation}
The expression ${\rm d} x^2 + {\rm d} y^2$, after the transformation
(\ref{10.13}) goes over into
\begin{equation}\label{10.23}
\left({\rm d} x_2^2 + {\rm d} y_2^2\right) / {\widetilde{W}_0}^2,
\end{equation}
and so the two equations above imply that after the transformation:
\begin{equation}\label{10.24}
\frac {{\rm d} x^2 + {\rm d} y^2} {{\cal E}^2} = \frac {{\rm d} {x_2}^2 + {\rm
d} {y_2}^2} {{S_0}^2 \left[(x_2 + \alpha)^2 + (y_2 + \beta)^2\right]^2}.
\end{equation}
We are in the same Szekeres model as at the beginning, but at a different
location. The alien form of the Szekeres metric in these coordinates results
from the fact that the transformation (\ref{10.13}) is not a symmetry, apart
from the single surface $z = z_0$. Since the quantities in this equation
($\alpha$, $\beta$ and $S_0$) depend on the continuous parameter $\lambda$, the
equation above describes the result of a shift to any location (with $\lambda =
0$ corresponding to identity). Equations (\ref{10.24}) and (\ref{10.21}) show
that the transformation (\ref{10.13}) is area-preserving.

The second integral in (\ref{10.5}) does not change its zero value under the
transformation (\ref{10.13}) -- since that transformation is simply a change of
variables under a definite integral, applied both to the integrand and to the
area of integration. Thus, the second integral in (\ref{10.5}) does not
contribute to calculating ${\cal M}$ in (\ref{10.3}).

 \section{Summary}

 \setcounter{equation} {0}

Continuing the research begun in Ref. \cite{HeKr2008}, geometrical properties of
the quasi-plane Szekeres model were investigated here, along with the
corresponding properties of the plane symmetric model. The following results
were achieved:

1. The pattern of decelerated expansion in the plane symmetric model was
analysed and shown to be in complete disagreement with the Newtonian analogues
and intuitions (Sections \ref{plsymmod} and \ref{newtplsym}). An example of a
Newtonian potential that gives a similar pattern of expansion has co-axial
parallel cones as its equipotential surfaces; it has not been investigated
whether such a potential can be generated by any realistic matter distribution.

2. Embeddings of the constant $t$ and constant $z$ surfaces in the Euclidean
space suggest that the flat surfaces contained in the plane symmetric model can
be interpreted as flat tori whose circumferences are proportional to the
function $R(t, z)$, and thus vary with time (Sec. \ref{PlanSym3Sp}). Such a
topology immediately explains the pattern of expansion and implies that the
total mass contained within a $z =$ const surface is finite.

3. The quasi-plane and quasi-hyperbolic models are permanently trapped (Sec.
\ref{noAH}), so no apparent horizons exist in them. Consequently, these models
cannot be used to describe the formation of black holes (the whole Universe is
one black hole all the time).

4. The quasi-plane model cannot describe the formation of structures that
collapse to very high densities (Sec. \ref{struform}), since the density
perturbations tend to finite values in the asymptotic future.

5. In the full (nonsymmetric) case the mass function is proportional to the
active gravitational mass within a 'wiggly tube' of finite radius (Sec.
\ref{infmassint}).

Whether the toroidal interpretation is a necessity is still unknown. However,
this paper demonstrated that with the toroidal topology the plane symmetric
model becomes in several respects simpler, and, however paradoxical this may
sound, more realistic.

The plane symmetric model with toroidal spaces may be a testing ground for the
idea of a 'small Universe', proposed by Ellis \cite{Elli1984}. A small Universe
is one with compact spatial sections, in which thus a present observer has
already seen several times around the space. Several papers were devoted to
checking this idea against the observational data (see, for example, Refs.
\cite{Lumi2003} -- \cite{Lach1995}; a conclusive proof or disproof of any
nontrivial topology is, unfortunately, still lacking). However, the background
geometry has always been a homogeneous isotropic Robertson -- Walker metric with
identifications in the underlying manifold. The plane symmetric toroidal
Szekeres (Ellis) model has a less general topology (identifications in it occur
only in two-dimensional surfaces, in the $z$-direction the space is infinite),
but is inhomogeneous, so might be useful for considering light propagation and
comparing the mass distribution in the model with the observed images.

{\bf Acknowledgements} The research for this paper was inspired by a
collaboration with Charles Hellaby, initiated in 2006 at the Department of
Mathematics and Applied Mathematics in Cape Town. It was supported by the Polish
Ministry of Science and Education grant no 1 P03B 075 29. Discussions with Ch.
Hellaby helped in clarifying several points. I am grateful to Dr. Stanis{\l}aw
Bajtlik for directing me to the relevant references on the observational aspects
of the topology of space. Members of the relativity seminar at the Institute of
Theoretical Physics, Warsaw University, are gratefully acknowledged for their
valuable comments on the geometry of projective planes that led to an
improvement in this text.

 \end{document}